\documentclass[aps,twocolumn]{revtex4}%
\usepackage{amsmath}
\usepackage{graphicx}
\usepackage{amsfonts}
\usepackage{epsfig}
\usepackage{amssymb}%
\begin{document}
\title[Non-Hermitian Hamiltonians with real eigenvalues coupled to electric fields]{Non-Hermitian Hamiltonians with real eigenvalues coupled to electric fields:\\from the time-independent to the time dependent quantum mechanical formulation}
\author{C. Figueira de Morisson Faria}
\author{A. Fring}
\affiliation{Centre for Mathematical Science, City University, Northampton Square, London
EC1V 0HB, UK}

\begin{abstract}
We provide a reviewlike introduction into the quantum mechanical formalism
related to non-Hermitian Hamiltonian systems with real eigenvalues. Starting
with the time-independent framework we explain how to determine an appropriate
domain of a non-Hermitian Hamiltonian and pay particular attention to the role
played by $\mathcal{PT}$-symmetry and pseudo-Hermiticity. We discuss the
time-evolution of such systems having in particular the question in mind of
how to couple consistently an electric field to pseudo-Hermitian Hamiltonians.
We illustrate the general formalism with three explicit examples: i) the
generalized Swanson Hamiltonians, which constitute non-Hermitian extensions of
anharmonic oscillators, ii) the spiked harmonic oscillator, which exhibits
explicit supersymmetry and iii) the $-x^{4}$-potential, which serves as a toy
model for the quantum field theoretical $\phi^{4}$-theory.

\end{abstract}
\volumeyear{year}
\volumenumber{number}
\issuenumber{number}
\eid{identifier}
\startpage{1}
\endpage{2}
\maketitle
\date[submitted]{July 2006}

\section{Introduction}

In general most physicists will almost instinctively associate a non-Hermitian
Hamiltonian with unstable states, decaying wavefunctions, resonances and
dissipation. Such type of systems have been studied for a long time. They
arise for instance when coupling channels in a system in which the
wavefunctions factorize into functions which depend on separate sets of
variables. The effective Hamiltonians resulting in this manner are
non-Hermitian and have \textit{complex} eigenvalues \cite{FW}. However, one
should note that Hermiticity of the Hamiltonian is only a sufficient
condition, which guarantees real eigenvalues and the conservation of
probability densities. It needs to be emphasized that it is not a necessary
condition and there could be non-Hermitian Hamiltonians with \textit{real}
discrete eigenvalue spectra, which then constitute potential candidates for
physical applications, such as for instance atomic systems without decay.

Precisely such type of Hamiltonian systems are currently under intense
investigation (for a collection of recent results see for instance
\cite{special}). The central question in this context is of course how to
obtain a consistent quantum mechanical framework. So far much effort has gone
into the study of time-independent eigenvalue problems. The main question we
wish to address here is how to couple an external time-dependent electric
field to a non-Hermitian Hamiltonian with real eigenvalues \cite{CA}.

Our manuscript is organized as follows: For the benefit of the non-expert and
the audience of this conference we commence in section II with a brief
reviewlike introduction by recalling some by now well-known facts and
arguments on the consistent quantum mechanical formulation of non-Hermitian
Hamiltonian systems. In section A we review the time-independent formulation
starting with a discussion of how to determine the appropriate domain for a
non-Hermitian Hamiltonian from the choice of asymptotic boundary condition. In
part 2 of this section we explain the limited role played by $\mathcal{PT}%
$-symmetry. In part 3 of section A we explain how pseudo-Hermiticity can be
employed to map almost all relevant problems in the non-Hermitian scenario to
a Hermitian system in the same equivalence class. The two systems obtained in
this manner are therefore isospectral. In section B we discuss how this
formalism can be extended to include an evolution in time. We describe here
gauge transformations, perturbation theory and how to compute various physical
quantities in the non-Hermitian setting. In section III we discuss two methods
of how to solve one of the key problems in this context, namely how to compute
pseudo-Hermitian Hamiltonians. Section IV contains three explicit examples to
which the formulation from the previous sections applies: i) the generalized
Swanson Hamiltonians, which constitute non-Hermitian extensions of anharmonic
oscillators, ii) the spiked harmonic oscillator, which exhibits explicit
supersymmetry and iii) the $-x^{4}$-potential, which serves as a toy model for
the quantum field theoretical $\phi^{4}$-theory. We state our conclusions and
an outlook to further problems in section V.

\section{The general framework}

\subsection{Time-independent quantum mechanical formulation}

\subsubsection{The domain of non-Hermitian Hamiltonians}

The current interest in this subject was triggered eight years ago
\cite{Bender:1998ke} by the at the time rather surprising numerical
observation that the Hamiltonian
\begin{equation}
H=p^{2}-g(iz)^{N} \label{1}%
\end{equation}
defined on a suitable domain possesses a real positive and discrete eigenvalue
spectrum for integers $N\geq2$ with positive real coupling constant $g$. This
property holds despite it being non-Hermitian $H\neq H^{\dagger}$ and
unbounded from below, for $N=4n$ with $n\in\mathbb{N}$ . Throughout this paper
we use atomic units $\hbar=e=m_{e}=c\alpha=1$.

Viewing now $H$ in (\ref{1}) as a differential operator in position space
acting on some wavefunction $\Phi(z)$, one needs to specify appropriate
boundary conditions in order to select a meaningful domain. In
\cite{Bender:1998ke} it was argued that the natural boundary condition,
$\Phi(z)\rightarrow0$ exponentially for $|z|\rightarrow\infty$, requires that
one continues the eigenvalue problem into the complex $z$-plane. In fact, for
$H$ in (\ref{1}) it was found that the wedges bounded by the Stokes lines in
which this boundary condition holds are given by%
\begin{align}
\mathcal{W}_{L}(N)  &  =\left\{  \theta\left\vert -\frac{8+N}{2(N+2)}%
\pi<\theta<-\frac{4+N}{2(N+2)}\pi\right.  \right\}  ,\label{We1}\\
\mathcal{W}_{R}(N)  &  =\left\{  \theta\left\vert -\frac{N}{2(N+2)}\pi
<\theta<\frac{4-N}{2(N+2)}\pi\right.  \right\}  , \label{We2}%
\end{align}
where $\theta=\arg z$. To see this one can follow the procedure for an
asymptotic expansion of a differential operator as outlined for instance in
\cite{BO}. Substituting $\Phi(z)=\exp(\varphi(z))$ into the eigenvalue
equation $H\Phi=\varepsilon\Phi$ yields $\varphi^{\prime\prime}+(\varphi
^{\prime})^{2}+g(iz)^{N}+\varepsilon=0$. For $\left\vert z\right\vert
\rightarrow\infty$ with infinity being an irregular singular point one may
assume that $\varphi^{\prime\prime}\ll(\varphi^{\prime})^{2}$. For large $z$
we can also neglect $\varepsilon$ in comparison with the potential and obtain
\begin{equation}
\varphi(z)\sim\frac{2\sqrt{g}}{N+2}i^{(1+\frac{N}{2})}z^{(1+\frac{N}{2}%
)}~~~\text{for~~ }\left\vert z\right\vert \rightarrow\infty.
\end{equation}
In order to extract the dominating exponential factor in $\Phi(z)$ and to
achieve $\Phi(z)\rightarrow0$ for $|z|\rightarrow\infty$, we require
$\operatorname{Re}\varphi(z)<0$. With $\theta=\arg z$ this is equivalent to%
\begin{equation}
\sin\left(  \frac{\pi N}{4}+\frac{2+N}{2}\theta\right)  >0, \label{sin}%
\end{equation}
which amounts to the conditions (\ref{We1}), (\ref{We2}) for the left and the
right wedge $\mathcal{W}_{L}$ and $\mathcal{W}_{R}$, respectively. Of course
(\ref{sin}) allows for many more solutions and therefore possible wedges, but
the selection criterion for (\ref{We1}), (\ref{We2}) is to reproduce the
conventional wedge for the harmonic oscillator for $N=2$, which is centered
around the real axis.

This means the domain of integration, which makes the eigenvalue problem of
the non-Hermitian Hamiltonian operator in (\ref{1}) in position space well
defined for the asymptotic boundary condition $\Phi(z)\rightarrow0$
exponentially for $|z|\rightarrow\infty$ is any path in the complex $z$-plane
which remains inside the wedges $\mathcal{W}_{L}$ and $\mathcal{W}_{R}$ when
it approaches complex infinity. This means any path parameterized as $z(x)$
with $x\in\mathbb{R}$, which satisfies%
\begin{equation}
\lim_{x\rightarrow\pm\infty}\arg[z(x)]\in\mathcal{W}_{R/L}%
\end{equation}
guarantees the appropriate boundary condition, namely exponential decay at
infinity of the wavefunction $\Phi(z)$. For various purposes, for instance
when one is concerned about a fast numerical convergence, one can also
determine the anti-Stokes lines, that is the domain on which the wavefunction
vanishes most rapidly, see e.g. \cite{BO}. For $H$ in (\ref{1}) the
anti-Stokes angles $\theta_{L/R}^{AS}$ are just in the centre of
$\mathcal{W}_{L}$ and $\mathcal{W}_{R}$ \cite{Bender:1998ke}.

Permissible domains are therefore usually some form of parameterizations for
hyperbolae. For instance a modified version of a parameterization used in
\cite{Mostafazadeh:2004tp} was suggested in \cite{JM1}
\begin{equation}
z_{1}(x)=x\cos(\theta_{R}^{AS})+i\sin(\theta_{R}^{AS})\sqrt{a^{2}+x^{2}%
}\label{z1}%
\end{equation}
with $a\in\mathbb{R}$. This clearly satisfied the required asymptotic%
\begin{equation}
\lim_{x\rightarrow\pm\infty}\arg[z_{1}(x)]=\theta_{R/L}^{AS}(N)\in
\mathcal{W}_{R/L}(N).
\end{equation}
for all values of $N$. However, as we shall discuss in more detail in section
IV certain manipulations depend crucially on the suitable choice of the
parameterization and one needs various alternatives. The selection procedure
for what is most "appropriate" is largely left to inspired guess work at this
stage. As we shall see below, an extremely useful variation of (\ref{z1}) was
provided in \cite{JM}%
\begin{equation}
z_{2}(x)=-2i\sqrt{1+ix}\label{z2}%
\end{equation}
with%
\begin{equation}%
\begin{array}
[c]{c}%
\lim\limits_{x\rightarrow\infty}\arg[z_{2}(x)]=-\frac{\pi}{4}\in
\mathcal{W}_{R}(N)\\
\lim\limits_{x\rightarrow-\infty}\arg[z_{2}(x)]=\frac{-3\pi}{4}\in
\mathcal{W}_{L}(N)
\end{array}
~
\end{equation}
for $N=3,4,\ldots,9$. Another permissible parameterization can be found for
instance\ \cite{Weide}. We illustrate the above discussion with some examples
in figure 1:

\noindent\epsfig{file=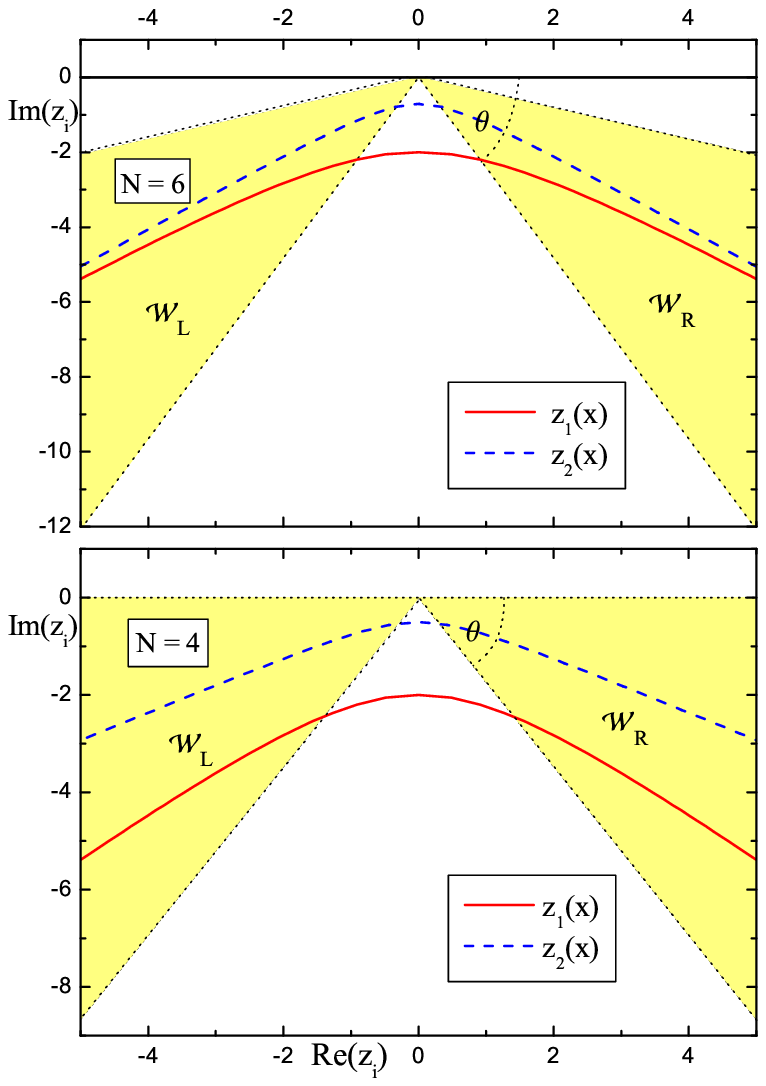,width=9.1cm}

{\small Figure 1: Stoke wedges in which the eigenfunctions of $H$ in (\ref{1})
for $N=4,6$ vanish exponentially when $\left\vert z\right\vert \rightarrow
\infty$. Permissible paths $z_{1}$ with a=1 and $z_{2}$ as parameterized in
(\ref{z1}) and (\ref{z2}), respectively. The Stokes lines are depicted as
dotted lines in the figure.}

\subsubsection{$\mathcal{PT}$-symmetry and real eigenvalues}

So how can one explain such unconventional behaviour that a non-Hermitian
Hamiltonian possesses a real eigenvalue spectrum? Shortly after the above
mentioned observation it was suggested that the reality of the spectrum should
be attributed to unbroken $\mathcal{PT}$-symmetry \cite{Bender:2002vv}, that
is the validity of the \textit{two} relations%
\begin{equation}
\left[  H,\mathcal{PT}\right]  =0\quad\text{and\quad}\mathcal{PT}\Phi=\Phi,
\label{PT}%
\end{equation}
where $\Phi$ is a square integrable eigenfunction on some domain of $H$. In
other words, when the Hamiltonian \textit{and} the wavefunction remain
invariant under a simultaneous parity transformation $\mathcal{P}$ and time
reversal $\mathcal{T}$%
\begin{equation}%
\begin{array}
[c]{llll}%
\mathcal{P}: & ~p\rightarrow-p\quad\quad & z\rightarrow-z & \\
\mathcal{T}: & ~p\rightarrow-p & z\rightarrow z & i\rightarrow-i\\
\mathcal{PT}: & ~p\rightarrow p & z\rightarrow-z\quad\quad & i\rightarrow-i,
\end{array}
\label{PTPT}%
\end{equation}
the eigenvalues of $H$ are real. As an example one sees that obviously the
Hamiltonian in equation (\ref{1}) is $\mathcal{PT}$-symmetric. What is less
straightforward to see is that for $N<2$ the second relation in (\ref{PT})
does not hold. Analytic arguments, which establish these facts for the
Hamiltonian (\ref{1}) may be found in \cite{DDT,Shin}.

We shall now outline to what extend $\mathcal{PT}$-symmetry can be utilized.
Clearly $\mathcal{P}^{2}=\mathcal{T}^{2}=(\mathcal{PT})^{2}=\mathbb{I}$ and
the last relation in (\ref{PTPT}) implies that the $\mathcal{PT}$-operator is
an anti-linear operator, i.e.~it acts as $\mathcal{PT}(\lambda\Phi+\mu
\Psi)=\lambda^{\ast}\mathcal{PT}\Phi+\mu^{\ast}\mathcal{PT}\Psi$ with
$\lambda,\mu\in\mathbb{C}$ and $\Phi,\Psi$ being eigenfunctions of the
Hamiltonian $H$ with eigenenergies $\varepsilon$, $H\Phi=\varepsilon\Phi$. The
anti-linear nature of the $\mathcal{PT}$-operator serves well to establish the
reality of the spectrum, i.e.~$\varepsilon=\varepsilon^{\ast}$, when
\textit{both} relations in (\ref{PT}) hold. This follows simply from%
\begin{align}
\varepsilon\Phi &  =H\Phi=H\mathcal{PT}\Phi=\mathcal{PT}H\Phi=\mathcal{PT}%
\varepsilon\Phi=\varepsilon^{\ast}\mathcal{PT}\Phi\nonumber\\
&  =\varepsilon^{\ast}\Phi. \label{44}%
\end{align}
Unfortunately, the anti-linearity is also responsible for the possibility that
only the first identity in (\ref{PT}) could hold, but not the second. In this
situation one speaks of a broken $\mathcal{PT}$-symmetry. The argument leading
to this is straightforward \cite{EW,SW}: Let us consider first a unitary
operator $U$ for which by definition%
\begin{equation}
\left\langle U\Psi\!\right.  \left\vert U\Phi\right\rangle =\left\langle
\Psi\!\right.  \left\vert \Phi\right\rangle \label{U}%
\end{equation}
holds for all eigenfunctions $\Phi,\Psi$ of $H$. From equation (\ref{U}) it
follows that $U\Psi=u\Psi$ with $\left\vert u\right\vert =1$ for all $\Psi$,
which means that a unitary operator has only one dimensional representations.
This property changes for anti-unitary operators $A$, as in that case only
$A^{2}$ is a unitary operator, which can be seen from%
\begin{equation}
\left\langle A^{2}\Psi\!\right.  \left\vert A^{2}\Phi\right\rangle
=\left\langle A\Phi\!\right.  \left\vert A\Psi\right\rangle =\left\langle
\Psi\!\right.  \left\vert \Phi\right\rangle . \label{A}%
\end{equation}
Now we can only deduce from (\ref{A}) that $A^{2}\Psi=a^{2}\Psi$ with
$\left\vert a^{2}\right\vert =1$ for all $\Psi$ and this means that an
anti-unitary (which is implied by anti-linearity) operator could have a two
dimensional representation $A\Psi=a^{\ast}\Phi$, $A\Phi=a\Psi$. Indeed when
$a$ is purely imaginary one can not construct a linear combination
$\Omega=\lambda\Phi+\mu\Psi,$ with $\lambda,\mu\in\mathbb{C}$ of the two
so-called flipping states $\Phi,\Psi$, which remains invariant under the
action of $A$. We see that%
\begin{equation}
A\Omega=\lambda^{\ast}a\Psi+\mu^{\ast}a^{\ast}\Phi=\lambda\Phi+\mu\Psi
\end{equation}
implies that $\mu=\lambda^{\ast}a$, $\lambda=\mu^{\ast}a^{\ast}$ and therefore
$a^{2}=1$. This means that only for $a=\pm1$ the two-dimensional
representation is reducible and for purely complex $a$ it is irreducible. In
the latter situation the second relation in (\ref{PT}) does therefore not
hold. From (\ref{44}) we see that $\mathcal{PT}\Phi$ is an eigenfunction of
$H$ with eigenvalue $\varepsilon^{\ast}$ when $\Phi$ is an eigenfunction of
$H$ with eigenvalue $\varepsilon$. Thus when the second relation in (\ref{PT})
does not hold the eigenvalues of $H$ come in complex conjugate pairs.

Thus $\mathcal{PT}$-symmetry is merely a fairly good guiding principle and
serves to identify immediately potentially interesting non-Hermitian
Hamiltonian systems. However, as argued above the $\mathcal{PT}$-symmetry of
$H$ does not constitute a guarantee for a real eigenvalue spectrum. It remains
an open question at this stage to determine under which circumstances the
$\mathcal{PT}$-symmetry is broken, albeit for Hamiltonians acting in a finite
dimensional Hilbert space an algorithm based on stability theory has been
provided \cite{brokenWeigert}. In addition one should stress, that
$\mathcal{PT}$-symmetry can not be regarded as the fundamental property, which
explains always the reality of the spectrum for non-Hermitian Hamiltonian
systems as there exist also examples with real spectra for which not even the
Hamiltonian is $\mathcal{PT}$-symmetric \cite{Ahmed,CA} (see also examples
below). In fact, more fundamental is the necessary and sufficient condition
that the Hamiltonian must be Hermitian with regard to \textit{some} positive
definite inner product \cite{Mostafazadeh:2001nr} as we shall discuss next.

We summarize the role played by $\mathcal{PT}$-symmetry in figure 2:

\noindent\epsfig{file=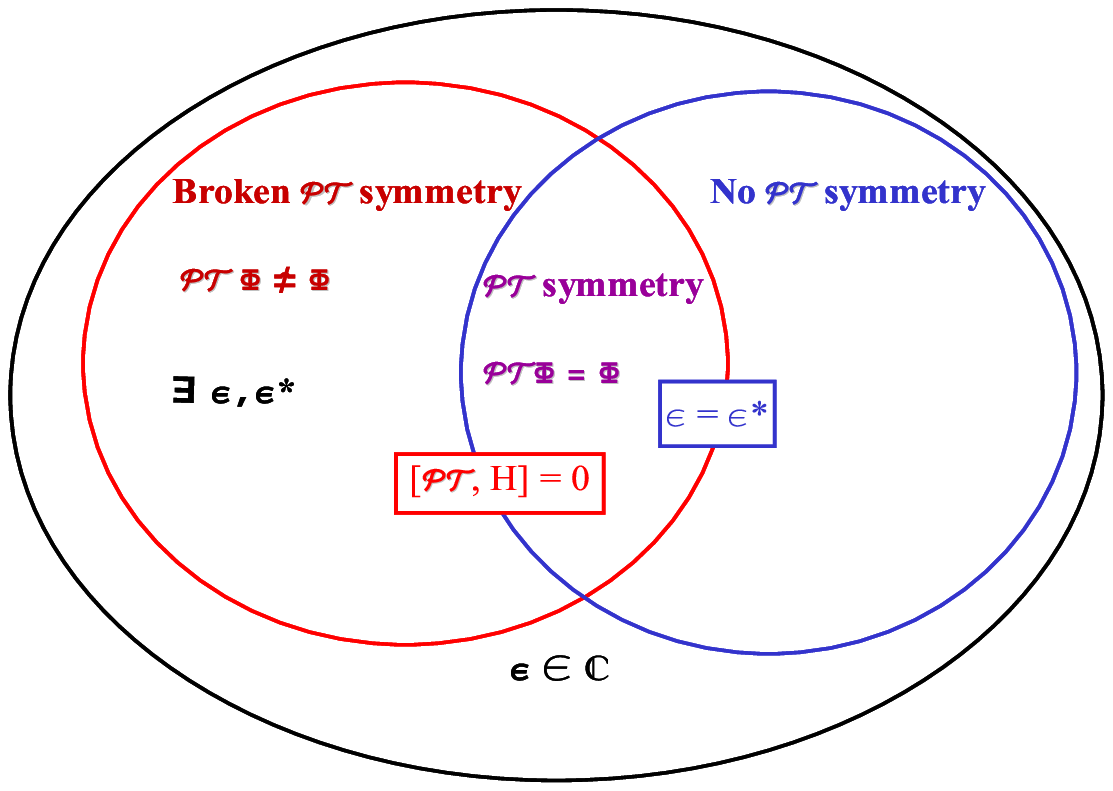,width=9.1cm}

{\small Figure 2: $\mathcal{PT}$-symmetry and real eigenvalues}

\subsubsection{Pseudo-Hermiticity and real eigenvalues}

The formal question of how to establish a consistent quantum mechanical
formalism for non-Hermitian Hamiltonian systems has already been discussed in
\cite{Urubu} prior to the above mentioned numerical observation. In fact the
possibility of extending a Hilbert space by a new intermediate state, which
then leads to an indefinite metric has already preoccupied particle physicists
more than half a century ago \cite{Heisenberg}. Some of these old results have
recently been re-discovered and developed further. As already mentioned
Hermiticity is a useful property as it guarantees the reality of the spectrum.
Let us briefly recall this standard argument and discuss how it needs to be
altered for the present scenario.

Suppose we have a diagonalizable Hermitian (symmetric) operator $h$ with
regard to the conventional inner product
\begin{equation}
\left\langle \phi_{n}\!\right.  \left\vert h\phi_{m}\right\rangle
=\left\langle h\phi_{n}\right\vert \!\left.  \phi_{m}\right\rangle
\text{.}\label{herm}%
\end{equation}
We use here Hermiticity in the sense that it implies self-adjointness and
ignore possible subtleties, which might arise from domain issues. In general
we understand here the domain to be the entire real axis. Multiplying next the
eigenvalue equations%
\begin{equation}
\left\vert h\phi_{m}\right\rangle =\varepsilon_{m}\left\vert \phi
_{m}\right\rangle \qquad\text{and\qquad}\left\langle h\phi_{n}\right\vert
=\varepsilon_{n}^{\ast}\left\langle \phi_{n}\right\vert \text{.}\label{S1}%
\end{equation}
by $\left\langle \phi_{n}\right\vert $ and $\left\vert \phi_{m}\right\rangle
$, respectively, we obtain%
\begin{align}
\left\langle \phi_{n}\!\right.  \left\vert h\phi_{m}\right\rangle  &
=\varepsilon_{m}\left\langle \phi_{n}\!\right.  \left\vert \phi_{m}%
\right\rangle \text{ }\label{3}\\
\left\langle h\phi_{n}\right\vert \!\left.  \phi_{m}\right\rangle  &
=\varepsilon_{n}^{\ast}\left\langle \phi_{n}\!\right.  \left\vert \phi
_{m}\right\rangle \text{.}\label{4}%
\end{align}
Taking the difference between (\ref{3}) and (\ref{4}) thus implies for $n=m$
that Hermiticity of $h$ with regard to the standard positive-definite inner
product $\left\langle \phi_{n}\right.  \left\vert \phi_{m}\right\rangle $,
i.e.~the validity of (\ref{herm}), is a sufficient condition for the energies
$\varepsilon_{n}$ to be real. Taking next $n\neq m$ Hermiticity then also
implies the orthogonality of the states $\left\vert \phi_{n}\right\rangle $
for all $n$.

It turns out that for non-Hermitian operators we only need to change the
definition for the inner product, i.e.~change the metric, to draw the same
conclusions \cite{Urubu,Mostafazadeh:2001nr}. Taking now the domains as
discussed in section 3, by definition we obviously no longer have
$\left\langle \Phi_{n}\right.  \!\left\vert H\Phi_{m}\right\rangle
=\left\langle H\Phi_{n}\!\right.  \left\vert \Phi_{m}\right\rangle $ for a
non-Hermitian operator $H$ with $\Phi_{n}$ obeying the eigenvalue equation
\begin{equation}
H\left\vert \Phi_{n}\right\rangle =\varepsilon_{n}\left\vert \Phi
_{n}\right\rangle . \label{S2}%
\end{equation}
Therefore there is no guarantee for the reality of the spectrum and neither
for the orthogonality. However, assuming $\eta$ to be a Hermitian operator
with respect to the standard inner product, we can define a new inner product
\begin{equation}
\left\langle \Phi_{n}\!\right.  \left\vert \Phi_{m}\right\rangle _{\eta
}:=\left\langle \Phi_{n}\!\right.  \left\vert \eta^{2}\Phi_{m}\right\rangle .
\label{eta}%
\end{equation}
Supposing now that $H$ is Hermitian with regard to this new inner product%
\begin{equation}
\left\langle \Phi_{n}\right.  \!\left\vert H\Phi_{m}\right\rangle _{\eta
}=\left\langle H\Phi_{n}\right.  \!\left\vert \Phi_{m}\right\rangle _{\eta},
\label{2}%
\end{equation}
we may employ exactly the same arguments as above and ensure the reality of
the spectrum as well as the orthogonality $\left\langle \Phi_{n}\!\right.
\left\vert \Phi_{m}\right\rangle _{\eta}=\delta_{n,m}$. Note that with regard
to the standard inner product one finds in general that $\left\langle \Phi
_{n}\!\right.  \left\vert \Phi_{m}\right\rangle \neq\delta_{n,m}$, see e.g.
\cite{Bender:2002vv}.

What is left is to characterize in more detail and possibly to determine is
the metric operator $\eta^{2}$. Mostafazadeh
\cite{Mostafazadeh:2002hb,Mostafazadeh:2001nr,Mostafazadeh:2002id,Mostafazadeh:2003gz}
proposed to assume that $H$ is a pseudo-Hermitian operator satisfying
\begin{equation}
h=\eta H\eta^{-1}=h^{\dagger}=\eta^{-1}H^{\dagger}\eta~~\Leftrightarrow
~~H^{\dagger}=\eta^{2}H\eta^{-2}, \label{sim}%
\end{equation}
where $\eta$ is the Hermitian operator with regard to the standard inner
product as introduced above. Since the Hermitian Hamiltonian $h$ and the
non-Hermitian Hamiltonian $H$ are related by a similarity transformation, they
belong to the same similarity class and therefore have the same eigenvalues.
The corresponding time-independent Schr\"{o}dinger equations are then simply
(\ref{S1}) and (\ref{S2}), where the corresponding wavefunctions are related
as
\begin{equation}
\Phi=\eta^{-1}\phi. \label{ff}%
\end{equation}
Having real eigenvalues for the Hermitian Hamiltonian $h$ then guarantees by
construction the same real eigenspectrum also for $H$. In fact the necessary
and sufficient condition (\ref{2}), which ensures the reality of the spectrum
for $H$ follows then from (\ref{herm})
\begin{align}
&  \left\langle \Psi\right.  \!\left\vert H\Phi\right\rangle _{\eta
}=\left\langle \Psi\right.  \!\left\vert \eta^{2}H\Phi\right\rangle
=\left\langle \eta^{-1}\psi\right.  \!\left\vert \eta^{2}H\eta^{-1}%
\phi\right\rangle \nonumber\\
&  =\left\langle \psi\right.  \!\left\vert \eta H\eta^{-1}\phi\right\rangle
=\left\langle \psi\right.  \!\left\vert h\phi\right\rangle =\left\langle
h\psi\right\vert \!\left.  \phi\right\rangle =\left\langle \eta H\eta^{-1}%
\psi\right\vert \!\left.  \phi\right\rangle \nonumber\\
&  =\left\langle H\Psi\right\vert \!\left.  \eta\phi\right\rangle
=\left\langle H\Psi\right\vert \!\left.  \eta^{2}\Phi\right\rangle
=\left\langle H\Psi\right\vert \!\left.  \Phi\right\rangle _{\eta}.
\end{align}

Clearly, when $\eta$ is Hermitian with regard to the standard inner product it
is also Hermitian with regard to the $\eta$-inner product (\ref{eta}).

A particular example for an $\eta$-inner product is the one introduced in
\cite{Bender:2002vv}
\begin{equation}
\left\langle \Psi\right\vert \!\left.  \Phi\right\rangle _{\mathcal{CPT}%
}:=\left(  \mathcal{CPT}\left\vert \Psi\right\rangle \right)  ^{T}%
\cdot\left\vert \Phi\right\rangle , \label{ICPT}%
\end{equation}
where the new operator $\mathcal{C}$, with $\mathcal{C}^{2}=\mathbb{I}$,
$[H,\mathcal{C}]=0$ and $[\mathcal{C},\mathcal{PT}]=0$, is employed. In
position space it reads $\mathcal{C}(x,y)=\sum\nolimits_{n}\Phi_{n}(x)\Phi
_{n}(y)$. The operators $\mathcal{C}$ and $\eta^{2}$ are simply related as
$\mathcal{C}=\eta^{-2}\mathcal{P}$.

In addition one should stress that in fact these inner products can also be
derived \cite{Mostafazadeh:2002pd,Weigertbi} when starting from a
biorthonormal basis, which is quite common to use in the study of
non-Hermitian Hamiltonian systems with complex eigenvalues, that is decaying
states, see e.g. \cite{Rot2,Ingrid}.

Crucial for a proper quantum mechanical framework is of course to clarify the
nature of the physical observables. In order to be suitable for a physical
interpretation, observables $\mathcal{O}$ have to be Hermitian operators
acting in some physical Hilbert space. From what has been outlined above with
regard to the inner products, it is natural to take them to be Hermitian with
respect to the new $\eta$-inner product%
\begin{equation}
\left\langle \Phi_{n}\right.  \!\left\vert \mathcal{O}\Phi_{m}\right\rangle
_{\eta}=\left\langle \mathcal{O}\Phi_{n}\right.  \!\left\vert \Phi
_{m}\right\rangle _{\eta}. \label{OH}%
\end{equation}
This implies immediately that when $o$ is an observable in the Hermitian
system, then
\begin{equation}
\mathcal{O}=\eta^{-1}o\eta\qquad\Leftrightarrow\qquad\mathcal{O}=\eta
^{-2}\mathcal{O}^{\dagger}\eta^{2} \label{obs}%
\end{equation}
is an observable in the non-Hermitian system. This means in turn that the
standard position operator $x$ and the momentum operators $p$ are in general
not observable in the non-Hermitian system, but rather their non-Hermitian
counterparts $X$ and $P$, respectively. Clearly $X$ and $P$ satisfy the
standard canonical commutation relations $\left[  X,P\right]  =i$ when
$\left[  x,p\right]  =i$. For Hamiltonians $h,H$, which admit a polynomial
expansion in $\{x,p\}$, $\{X,P\}$, it follows then directly from (\ref{sim})
that
\begin{align}
H(x,p)  &  =\eta^{-1}h(x,p)\eta=h(X,P),\label{89}\\
h(x,p)  &  =\eta^{-1}H^{\dagger}(x,p)\eta=H^{\dagger}(X,P). \label{90}%
\end{align}
These relations serve for instance as a consistency check when we start with a
given non-Hermitian Hamiltonian and construct its Hermitian counterpart by
means of a similarity transformation. Moreover (\ref{89}) provides a simple
way to express the non-Hermitian Hamiltonian in terms of the canonical
$(X,P)$-variables, which have a physical meaning for that system rather than
the $(x,p)$-variables, which are in general meaningless in that context. In
addition, one may use (\ref{89}) as a principle to construct non-Hermitian
Hamiltonians with real spectra from a given Hermitian Hamiltonian and a set of
canonical variables and vice versa with (\ref{90}). When $\eta$ is
$\mathcal{PT}$-symmetric in the $(x,p)$-variables the corresponding quantities
in the non-Hermitian system will be $\mathcal{PT}$-symmetric in the $(X,P)$-variables.

We conclude with a final comment in regard to the uniqueness of the metric
operator $\eta^{2}$. In fact there are various types of ambiguities arising,
which we comment on in section III when we discuss how to compute $\eta^{2}$
explicitly. In \cite{Urubu} Scholtz, Geyer and Hahne proved that the metric
operator $\eta^{2}$ is uniquely determined on a Hilbert space if and only if a
set of observables $\mathcal{O}_{i}$ with respect to (\ref{OH}) is irreducible
on this Hilbert space. The latter means that there is no bounded operator
besides the identity, which commutes with all observables $\mathcal{O}_{i}$.
Taking this result into account allows to move the nature of the ambiguities
from the metric to the specification of the set of observables. As we shall
see below a subset or even one observable might be enough in practise.

\subsection{Time-dependent quantum mechanical formulation}

Let us now discuss how to couple a laser field to the non-Hermitian
Hamiltonians $H$, which have the properties described above. In the simplest
scenario, i.e.~if the parameters involved lie within a non-relativistic
regime and the dipole approximation holds, such a field can be approximated by
a time-dependent electric field $E(t)$. In the following, we will briefly
recall our recent results on the temporal evolution of the resulting system
\cite{CA}. For simplicity, we assume that $E(t)$ is linearly polarized and has
a finite duration $\tau$.

Within a Hermitian framework and in the length gauge, such an evolution is
described by the\ time-dependent Schr\"{o}dinger equation
\begin{equation}
i\partial_{t}\phi(t)=h_{l}(t)\phi(t), \label{hl}%
\end{equation}
where
\begin{equation}
h_{l}(t)=\frac{p^{2}}{2}+V(x)+xE(t)=h+xE(t) \label{20}%
\end{equation}
is the Stark-LoSurdo Hamiltonian \cite{LPR,HS}. For a pulse of finite
duration, $h\phi(0)=\varepsilon\phi(0)$ and $h\phi(\tau)=\varepsilon\phi
(\tau)$. We assume here that $h$ possesses a non-Hermitian counterpart $H$
which is in the same equivalence class, i.e.~the validity of the first
relation in (\ref{sim}).

\subsubsection{Time-evolution operators}

The central quantity of interest in this context is the time-evolution
operator
\begin{equation}
u(t,t^{\prime})=T\exp\left(  -i\int\nolimits_{t^{\prime}}^{t}dsh(s)\right)  ,
\label{timeev}%
\end{equation}
which evolves a wavefunction from a time $t^{\prime}$ to $t$, that is
$\phi(t)=u(t,t^{\prime})\phi(t^{\prime})$. In (\ref{timeev}), $T$ denotes the
time ordering. One should note that, in general, $u(t,t^{\prime})\neq
\exp\left[  -ih(t-t^{\prime})\right]  .$ In fact, such a relation only holds
for Hamiltonians which are not explicitly time-dependent, as is not the case
for the scenario we have in mind.

When $h(s)$ is a self-adjoint operator in some Hilbert space, $u(t,t^{\prime
})$ satisfies the relations \cite{AC1,AC2,AC3}
\begin{align}
i\partial_{t}u(t,t^{\prime})  &  =h(t)u(t,t^{\prime}),\ \nonumber\\
u(t,t^{\prime})u(t^{\prime},t^{\prime\prime})  &  =u(t,t^{\prime\prime
})\ \text{and\ }u(t,t)=\mathbb{I~}. \label{u}%
\end{align}

We will now assume that the similarity transformation $\eta$ extends to the
time-dependent case. Thus, $H(t)=\eta^{-1}h(t)\eta$ , with $H(t)\neq
H^{\dagger}(t).$ We take $\eta$ to be time-independent. This allows us to
guarantee that the relations
\begin{align}
i\partial_{t}U(t,t^{\prime})  &  =H(t)U(t,t^{\prime}),\nonumber\\
\ U(t,t^{\prime})U(t^{\prime},t^{\prime\prime})  &  =U(t,t^{\prime\prime
})\ \text{and\ }U(t,t)=\mathbb{I},\mathbb{~} \label{nhU}%
\end{align}
for the time-evolution operator $U(t,t^{\prime})=\eta^{-1}u(t,t^{\prime})\eta$
associated to the non-Hermitian Hamiltonian $H(t),$ also hold. Then this
operator fulfills the condition $U^{\dagger}(t,t^{\prime})=\eta^{2}%
U^{-1}(t,t^{\prime})\eta^{-2},$ which follows from $u^{\dagger}(t,t^{\prime
})=u^{-1}(t,t^{\prime})$.

It is worth stressing that we make no simplifying assumption on the time
dependence of the Hamiltonian. In fact, the only requirements involved in our
approach are i) that $H$ be pseudo-Hermitian and ii) that the similarity
transformation $\eta$ be time-independent. Such conditions guarantee that the
time-dependent Schr\"{o}dinger equation and the relations involving the time
evolution operator remain valid also in the non-Hermitian case.

These conditions, however, are far more general than those normally
encountered in the literature. In fact, most studies make several simplifying
assumptions on the time-evolution operator, in the sense that they concentrate
on Hamiltonians which are either not explicitly time-dependent, or which vary
adiabatically and/or periodically with time. The first scenario is addressed
by either solving the eigenvalue problem $H\Phi=\varepsilon\Phi,$ or, at most,
by employing the time-evolution operators $U(t,t^{\prime})=\exp
[-iH(t-t^{\prime})]$.

The remaining situations are widespread in the atomic physics literature, in
the context of open quantum systems. Roughly speaking, if a system is close to
the adiabatic limit this means that it is varying so slowly that the problem
can be reduced to solving eigenvalue equations of the form $H(t)\Phi
(t)=\varepsilon_{n}(t)\Phi(t)$. In a standard, Hermitian framework, this
implies that $\partial_{t}u(t)u^{\dagger}(t)\ll u(t)h(t)u^{\dagger}(t)$, and
that transitions between different time-dependent eigenstates of $H(t)$ will
be induced by perturbations around the adiabatic limit. Specifically for a
system coupled to an external laser field, the time-dependent energies
$\varepsilon_{n}(t)$ give the field-dressed states (for a first derivation of
the adiabatic theorem and for an extension of such a theorem to non-Hermitian
open quantum systems, see \cite{Born} and \cite{timedep5}, respectively). For
periodic fields, such a procedure is closely related to the Floquet theory,
for which there also exists time-dependent \textquotedblleft
quasienergies\textquotedblright. This approach may be problematic if the field
varies abruptly with time, such as, for instance, if it is an ultrashort pulse.

\subsubsection{Time-dependent physical quantities}

The time-evolution operators characterized in the previous subsection may then
be employed to compute various quantities of physical interest, such as for
instance the transition probability
\begin{equation}
\mathcal{P}_{n\leftarrow m}=\!\left\vert \left\langle \Phi_{n}\right\vert
U(t,0)\left.  \Phi_{m}\right\rangle _{\eta}\right\vert ^{2}\!\!=\left\vert
\left\langle \phi_{n}\right\vert u(t,0)\left.  \phi_{m}\right\rangle
\right\vert ^{2}\!\!, \label{P}%
\end{equation}
from an eigenstate $\left\vert \phi_{m}\right\rangle $ to $\left\vert \phi
_{n}\right\rangle $ of the Hermitian electric field-free Hamiltonian $h$ or
eigenstate $\left\vert \Phi_{m}\right\rangle $ to $\left\vert \Phi
_{n}\right\rangle $ of the non-Hermitian electric field-free Hamiltonian $H$.
Another physical quantity of interest is the time evolution for the
expectation value of an observable in the state $n$ is
\begin{align}
\mathcal{O}_{n}(t)  &  =\left\langle U(t,0)\Phi_{n}(0)\right\vert \!\!\left.
\mathcal{O}U(t,0)\Phi_{n}(0)\right\rangle _{\eta}\nonumber\\
&  =\left\langle u(t,0)\phi_{n}(0)\right\vert \!\!\left.  ou(t,0)\phi
_{n}(0)\right\rangle _{\eta}\\
&  =o_{n}(t).\nonumber
\end{align}
In a similar way we may proceed to compute ionization rates and probabilities
etc., but these examples are sufficient to see that, as in the
time-independent scenario, the relevant computations for the non-Hermitian
system can be translated into the Hermitian one, provided the $\eta$-operator
is known.

\subsubsection{Gauge transformations}

Apart from employing the length-gauge Hamiltonian $h_{l}(t)$, one may describe
a Hermitian Hamiltonian system coupled to an electric field in other gauges.
Concrete examples are the velocity gauge, obtained by employing the
minimal-coupling prescription $p\rightarrow p-b(t)$, or the
Kramers-Henneberger gauge, obtained with the shift $x\rightarrow x-c(t)$ in
the field-free Hamiltonian $h$ as introduced in (\ref{20}). The corresponding
Hamiltonians are given by
\begin{equation}
h_{v}(t)=\frac{(p-b(t))^{2}}{2}+V(x)=h(p-b(t)) \label{hv}%
\end{equation}
and%
\begin{equation}
h_{KH}(t)=\frac{p^{2}}{2}+V(x-c(t))=h(x-c(t)), \label{hKH}%
\end{equation}
respectively. In equation (\ref{hv}) and (\ref{hKH}),
\begin{equation}
b(t)=\int\nolimits_{0}^{t}dsE(s),\qquad c(t)=\int\nolimits_{0}^{t}dsb(s)\quad
\end{equation}
are the momentum transfer $b(t)$ from the laser field to the system in
question and the classical displacement $c(t)$ in the system caused by the
laser field.

Depending on the problem at hand, the gauge choice may considerably facilitate
the computations. For instance, the length gauge is very appropriate for
perturbation theory in the electric field, as the field coupling involves only
one additional term, or for physical interpretations in the low-frequency
regime, since it allows the physical picture of an effective time-dependent
potential. The Kramers-Henneberger gauge is most useful in the high-frequency
regime, especially if one wishes to exploit the periodicity of the field and
perform Floquet expansions. Each formulation can be obtained from the other
employing gauge transformations. The Hamiltonians in the length, velocity and
Kramers-Henneberger gauge are related by
\begin{equation}
h_{l}(p,x)-xE(t)=h_{v}(p+b(t),x)=h_{KH}(p,x+c(t)). \label{gaugeherm}%
\end{equation}

We will now perform such transformations for non-Hermitian Hamiltonian
systems. First, we will replace the wavefunction $\phi$ in the time-dependent
Schr\"{o}dinger equation related to the Hamiltonian $h$ by $\phi=a(t)^{-1}%
\phi^{\prime}$, with $a(t)$ being some unitary operator. This yields
\cite{AC1,AC2,AC3}%
\begin{equation}
i\partial_{t}\phi^{\prime}\!=\!h^{\prime}(t)\phi^{\prime}\!\!=\!\!\left[
a(t)h(t)a(t)^{-1}+i\partial_{t}a(t)a(t)^{-1}\right]  \phi^{\prime}.
\end{equation}
Specifically, the standard transformation from the length to the velocity
gauge, and from the velocity to the Kramers-Henneberger gauge, which are
extensively used in strong-field laser physics, are given by,
\begin{equation}
a_{l\rightarrow v}(t)=e^{ib(t)x}~~~\text{and~~~}a_{v\rightarrow KH}%
(t)=e^{id(t)}e^{-ic(t)p}. \label{standgauge}%
\end{equation}
respectively. In equation (\ref{standgauge}), in addition to the momentum
transfer and classical displacement, we have also introduced the classical
energy transfer $d(t)=\frac{1}{2}\int\nolimits_{0}^{t}dsb(s)^{2}$.

If the system is pseudo-Hermitian, one may employ the relation $\phi=\eta\Phi$
to obtain the gauge transformation%
\begin{equation}
i\partial_{t}\Phi^{\prime}\!\!=\!\!\left[  A(t)H(t)A(t)^{-1}+\!i\partial
_{t}A(t)A(t)^{-1}\right]  \Phi^{\prime}, \label{transfo}%
\end{equation}
where
\begin{equation}
a(t)=\eta A(t)\eta^{-1}\ \text{and\ }h(t)=\eta H(t)\eta^{-1},
\end{equation}
and the expression in brackets, on the right-hand-side of (\ref{transfo}),
denotes the gauge-transformed Hamiltonian $H^{\prime}(t)$. The gauge
transformations $A(t)$, as it should be, guarantee the invariance of the
physical observables, when computed using the generalized inner product
(\ref{eta}). Now the relations
\begin{align}
H_{l}(X,P)  &  =H_{v}(X,P+b(t))+XE(t)\label{gaugenonherm}\\
&  =H_{KH}(X+c(t),P)+XE(t),\nonumber
\end{align}
hold for pseudo-Hermitian Hamiltonians.

\subsubsection{Perturbation theory}

Since, in most realistic situations, the time-dependent Schr\"{o}dinger
equation describing the evolution of a system with a binding potential $V(x)$
subjected to a time-dependent laser field $E(t)$ does not possess an analytic
solution, it is necessary to resort to perturbation theory. In order to
construct a perturbative series in a pseudo-Hermitian framework, we will
initially consider a time-dependent Hermitian Hamiltonian $h(t)=h_{0}%
(t)+h_{p}(t)$, where $h_{0}(t)$ and $h_{p}(t)$ are also Hermitian and satisfy
the time-dependent Schr\"{o}dinger equation. Using the Du Hamel formula
\cite{AC1,AC2,AC3}, we can express the time-evolution operator$\ u(t,t^{\prime
})$ associated to $h(t)$ as
\begin{equation}
u(t,t^{\prime})=u_{0}(t,t^{\prime})-i\int\nolimits_{t^{\prime}}^{t}%
u(t,s)h_{p}(s)u_{0}(s,t^{\prime})ds, \label{DuHamel1}%
\end{equation}
where $u_{0}(t,t^{\prime})$ is the time evolution operator with respect to
$h_{0}(t)$. Equation (\ref{DuHamel1}) can then be solved iteratively to an
arbitrary order in $h_{p}(t),$ which will be the perturbation. Roughly
speaking if \ $h_{p}(t)\ll h_{0}(t)$, the series obtained by such means has a
great chance to converge. For instance, for weak laser fields and in the
length gauge, a natural choice is to take $h_{p}(t)=xE(t)$ and $h_{0}%
(t)=p^{2}/2+V$, whereas in the strong-field regime we take $h_{0}%
(t)=p^{2}/2+xE(t)$ as the Gordon-Volkov Hamiltonian and the perturbation is
chosen as $h_{p}(t)=V.$

Similarly, for the time evolution operator $U(t,t^{\prime})$ related to its
pseudo-Hermitian counterpart $H(t)=H_{0}(t)+H_{p}(t)$, with $H_{0}%
(t)=\eta^{-1}h_{0}(t)\eta$ and $H_{p}(t)=\eta^{-1}h_{p}(t)\eta$, we may also
write
\begin{equation}
U(t,t^{\prime})=U_{0}(t,t^{\prime})-i\int\nolimits_{t^{\prime}}^{t}%
U(t,s)H_{p}(s)U_{0}(s,t^{\prime})ds, \label{DuHamel2}%
\end{equation}
where $U_{0}(t,t^{\prime})$ is related to the Hamiltonian $H_{0}(t)$, and the
perturbative series is obtained by iterating equation (\ref{DuHamel2}) up to
the desired order.\qquad

\paragraph{The weak intensity regime}

As argued in the previous subsection one can in general not compute the
time-evolution operator exactly and has to resort to perturbation theory
instead. We illustrate here briefly how this works more explicitly in the
different intensity regimes.

We commence with the weak intensity regime and we will consider first-order
perturbation theory with respect to the external laser field amplitude $E_{0}%
$. Iterating (\ref{DuHamel2}) it follows that to this order the time-evolution
operator can be approximated by
\begin{equation}
U^{(1)}(t,0)=U_{0}(t,0)-i\int\limits_{0}^{t}U_{0}(t,s)XE(s)U_{0}(s,0)ds,
\end{equation}
where $U_{0}(t,0)=\exp[-iHt].$ Subsequently the transition probability
(\ref{P}) from a state $m$ to $n$ to this order becomes
\begin{equation}
\mathcal{P}_{n\leftarrow m}\!=\!\left\vert \delta_{nm}-i\!\left\langle
\Phi_{n}\right\vert \!\!\left.  X\Phi_{m}\right\rangle _{\eta}\!\int
\limits_{0}^{t}\!dse^{i(\varepsilon_{n}-\varepsilon_{m})s}E(s)\right\vert
^{2}\!\!. \label{u1}%
\end{equation}
Note here the occurrence of the matrix element $\left\langle \Phi
_{n}\right\vert \!\!\left.  X\Phi_{m}\right\rangle _{\eta}=\left\langle
\phi_{n}\right\vert \left.  x\phi_{m}\right\rangle ,$ which results from
taking the non-Hermitian version of the Stark-LoSurdo Hamiltonian in
(\ref{20}) to be $H_{l}(t)=H+XE(t)$. In case we add $xE(t)$ instead of $XE(t)$
the amplitude $\left\langle \phi_{n}\right\vert \eta x\eta^{-1}\left.
\phi_{m}\right\rangle $ would occur. With our examples below we demonstrate
that the latter matrix element is very often unphysical.

\paragraph{The strong field regime}

Next we will address the opposite scenario, namely the situation in which the
laser field is larger, or at least comparable to the atomic binding forces.
Such a physical framework has become of interest since the mid-1980's, when
intense lasers became feasible, due to the wide range of phenomena and
applications existing in this context. Concrete examples are high-order
harmonic generation, above-threshold ionization, or laser-induced single and
multiple ionization (for reviews we refer to \cite{HH3,ATI,strong}). In this
case, it is a common procedure to perturb around the Gordon-Volkov
Hamiltonian, which, in a non-Hermitian framework and in the length gauge, is
given by $H_{l}^{(GV)}(t)=P^{2}/2+XE(t)$. To first order, the time-evolution
operator then reads
\begin{align}
U^{(1)}(t,0) &  =U_{l}^{(GV)}(t,0)\label{strongU}\\
&  -i\int\limits_{0}^{t}U_{l}^{(GV)}(t,s)V(X)U_{l}^{(GV)}(s,0)ds,\nonumber
\end{align}
where the Gordon-Volkov time-evolution operator is given by%
\begin{equation}
U_{l}^{(GV)}(t,0)=A_{KH\rightarrow l}(t)\exp[-iP^{2}t/2]A_{KH\rightarrow
l}^{-1}(0).
\end{equation}
The gauge transformation $A_{KH\rightarrow l}(t)$, from the Kramers
Henneberger to the length gauge, is written as%
\begin{align}
A_{KH\rightarrow l}(t) &  =\eta^{-1}e^{ic(t)p}e^{id(t)}e^{-ib(t)x}\eta\\
&  =e^{ic(t)P}e^{id(t)}e^{-ib(t)X}.
\end{align}
Obviously, one may also define a Gordon-Volkov Hamiltonian in the velocity
gauge as $H_{l}^{(GV)}(t)=(P-b(t))^{2}/2.$ In this case, the corresponding
time evolution operator is $U_{v}^{(GV)}(t,0)=e^{ib(t)X}U_{l}^{(GV)}%
(t,0)e^{-ib(0)X}.$

\section{Computing Pseudo-Hermitian Hamiltonians}

Having discussed the central role played by pseudo-Hermitian Hamiltonians it
is vital to have a constructive method to realize them. In other words we wish
to compute Hamiltonians $h=h^{\dagger}$ and $H\neq H^{\dagger}$ belonging to
the same equivalence class. This is a well defined problem, but in most cases
very difficult to solve. Here we present two different types of methods to
achieve this.

\subsection{Similarity transformations from operator identities}

Supposing that the similarity transformation (\ref{sim}) can be realized using
a Hermitian operator of the form $\eta=\exp(q/2)$, the second relation in
(\ref{sim}) implies by standard Baker-Campbell-Hausdorff commutation relations
that
\begin{align}
H^{\dagger}  &  =H+\left[  q,H\right]  +\frac{1}{2!}\left[  q,\left[
q,H\right]  \right]  +\frac{1}{3!}\left[  q,\left[  q,\left[  q,H\right]
\right]  \right]  +\ldots\nonumber\\
&  =\sum\limits_{n=0}^{\infty}\frac{1}{n!}c_{q}^{(n)}(H). \label{bkh}%
\end{align}
For convenience we have introduced here a more compact notation for the
$n$-fold commutator of the operator $q$ with some arbitrary operator
$\mathcal{O}$ as
\begin{equation}
c_{q}^{(n)}(\mathcal{O}):=\left[  q,\left[  q,\left[  q,\ldots\left[
q,\mathcal{O}\right]  \ldots\right]  \right]  \right]  .
\end{equation}
Clearly, if for some integer $n$ the $n$-fold commutator $c_{q}^{(n)}(H)$
vanishes the conjugation and therefore the similarity transformation can be
computed exactly. In order to see this more explicitly we separate next the
non-Hermitian Hamiltonian into its real and imaginary part and bring it into
the form
\begin{equation}
H=h_{0}+ih_{1},
\end{equation}
with $h_{0}=h_{0}^{\dagger}$, $h_{1}=h_{1}^{\dagger}$. For the case when one
has the condition $c_{q}^{(\ell+1)}(h_{0})=0$ for some finite integer $\ell$,
we found in \cite{CA} the closed expressions
\begin{align}
h  &  =h_{0}+\sum\limits_{n=1}^{[\frac{\ell}{2}]}\frac{(-1)^{n}E_{n}}%
{4^{n}(2n)!}c_{q}^{(2n)}(h_{0}),\quad\label{HHH}\\
H  &  =h_{0}-\sum\limits_{n=1}^{[\frac{\ell+1}{2}]}\frac{\kappa_{2n-1}%
}{(2n-1)!}c_{q}^{(2n-1)}(h_{0}), \label{HH}%
\end{align}
which are related according to the first identity in (\ref{sim}). Here
$\left[  x\right]  $ denotes the integer part of a number $x$. The $E_{n}$ are
Euler's numbers
\begin{equation}
E_{1}=1,\quad E_{2}=5,\quad E_{3}=61,\quad E_{4}=1385,\ldots
\end{equation}
and the $\kappa_{2n-1}$ may be computed from them according to%
\begin{equation}
\kappa_{n}=\frac{1}{2^{n}}\sum\limits_{m=1}^{\left[  (n+1)/2\right]
}(-1)^{n+m}\binom{n}{2m}E_{m}. \label{kan}%
\end{equation}
The first examples are
\begin{equation}
\kappa_{1}=\frac{1}{2},\quad\kappa_{3}=-\frac{1}{4},\quad\kappa_{5}=\frac
{1}{2},\quad\kappa_{7}=-\frac{17}{8},\ldots\label{ks}%
\end{equation}

Depending on how large $\ell$ becomes the explicit evaluation of sums in
(\ref{HHH}) and (\ref{HH}) can become rather complicated. In fact, in most
cases the series does not terminate and one has to compute the expressions
perturbatively. We shall not discuss such cases here and refer instead to the
literature \cite{Bender:2004sa,Mostafazadeh:2004qh,CA,ACIso,Can}.

\subsection{Similarity transformations from differential equations}

Alternatively one can follow a proposal put forward by Scholtz and Geyer
\cite{Moyal1,Moyal2} and solve (\ref{sim}) by means of Moyal products instead
of computing commutators. The central idea is to exploit isomorphic relations
between commutator relations and real valued functions multiplied by Moyal
products, which correspond to differential equations. We shall demonstrate
that this approach is rather practical and allows to compute pairs of
isospectral Hamiltonians $h=h^{\dagger}$ and $H\neq H^{\dagger}$, when they
are of polynomial nature.

We use a slightly different definition for the Moyal product as in
\cite{Moyal1,Moyal2}, since then the resulting differential equations become
simpler \cite{ACIso}. Following for instance \cite{Fairlie:1998rf} we define
the Moyal product of real valued functions depending on the variables $x$ and
$p$ as
\begin{align}
&  f(x,p)\star g(x,p)=f(x,p)e^{\frac{i}{2}(\overleftarrow{{\partial}}%
_{\!\!x}\overrightarrow{{\partial}}_{\!\!p}-\overleftarrow{{\partial}}%
_{\!\!p}\overrightarrow{{\partial}}_{\!\!x})}g(x,p)=\label{Moy}\\
&  \sum\limits_{s=0}^{\infty}\frac{(\frac{-i}{2})^{s}}{s!}\sum\limits_{t=0}%
^{s}(-1)^{t}\left(
\begin{array}
[c]{r}%
s\\
t
\end{array}
\right)  \partial_{x}^{t}\partial_{p}^{s-t}f(x,p)\partial_{x}^{s-t}%
\partial_{p}^{t}g(x,p).\nonumber
\end{align}
One may then use this expression to turn all operator identities into
differential equations. In principle this yields differential equations of
infinite order, but when $f(x,p),g(x,p)$ are of polynomial nature the series
terminates and the order will be finite. For instance, if we want to compute
the commutator $\left[  \hat{x},\hat{p}\right]  =i$ we have to evaluate the
corresponding Moyal product relation $x\star p-p\star x=i$. Here and in some
places below we emphasize the operator nature of the quantities involved by
dressing them with hats. In order to keep notations simple we do not always
make this rigorous distinction, when it is not strictly necessary. Matters
become more complicated when the resulting real valued function depends on $x$
as well as on $p$. As for a function the ordering is of course irrelevant we
need a prescription of how to turn such a function back into operator valued
expressions. Computing for instance
\begin{equation}
\lbrack\hat{x}^{2},\hat{p}^{2}]=4i\hat{p}\hat{x}-2~~~\cong~~x^{2}\star
p^{2}-p^{2}\star x^{2}=4ipx,
\end{equation}
we observe that we obtain the correct operator valued expression for the last
equality when we replace $px\rightarrow(px+xp)/2$. In general we have to
replace each monomial $p^{m}x^{n}$ or $x^{n}p^{m}$ by the totally symmetric
polynomial $S_{m,n}$ in the $m$ operators $p$ and $n$ operators $x$%
\begin{equation}
S_{m,n}=\frac{m!n!}{(m+n)!}\sum\limits_{\pi}p^{m}x^{n}. \label{Snm}%
\end{equation}
The sum extends over the entire permutation group $\pi$. For our purposes we
have usually a given non-Hermitian Hamiltonian $H$ and wish to compute from
the second relation in (\ref{sim}) the Hermitian operator $\eta^{2}.$ The
corresponding differential equation is then simply
\begin{equation}
H^{\dagger}(x,p)\star\eta^{2}(x,p)=\eta^{2}(x,p)\star H(x,p). \label{He}%
\end{equation}
Subsequently, one may compute also $\eta(x,p)$ and $h(x,p)$ in a similar manner.

A comment is due concerning the uniqueness of the solutions. Having solved
various differential equations, we naturally expect some ambiguities in the
general solutions, which mirror the possibility of different boundary
conditions. However, one should emphasize that these ambiguities are not only
present when using Moyal products, but are a general feature occurring also
when using commutation relations of the type (\ref{HHH}) and (\ref{HH}). It is
clear that in that context one may only fix the operator $q$ up to any
operator which commutes with the Hermitian part of $H$, that is $h_{0}$. This
means that, in (\ref{HHH}) and (\ref{HH}), the expressions are insensitive to
any replacement $q\rightarrow q+\tilde{q}$ with $[\tilde{q},h_{0}]=0$. A
further type of ambiguity, which is always present is a multiplication of
$\eta^{2}$ by operators which commute with $H$, i.e.~we could re-define
$\eta^{2}\rightarrow\eta^{2}Q$ for any $Q$, which satisfies $[Q,H]=0$.

It should be mentioned that there are also other possibilities to evaluate the
similarity transformations, such as for instance suggested in \cite{MOT} or
directly by using properties of differential equations \cite{BBC}.

Let us now demonstrate with some concrete examples how the above mentioned
formalism can be applied.

\section{(Quasi) exactly solvable models}

Non-Hermitian Hamiltonians may arise for various different reasons. In the
following we provide three such examples, which all arise from quite different
argumentations and thus provide several types of motivations to study
non-Hermitian Hamiltonian systems.

\subsection{The generalized Swanson Hamiltonian}

One type of non-Hermitian Hamiltonian system arises form a purely mathematical
consideration simply by perturbing a Hermitian Hamiltonian by adding a
non-Hermitian term. We start with a straightforward example, which results
when perturbing the anharmonic oscillators%
\begin{equation}
h_{n}^{0}(\alpha)=\frac{1}{2}p^{2}+\frac{\alpha}{2}x^{n}%
\end{equation}
for $n=1,2,3,\ldots$ and $\alpha\in\mathbb{R}$. Defining now the Hermitian
operators $\eta_{m}=\exp(q_{m}/2)$ with $q_{m}=2g/mx^{m}$ for $m=1,2,3,\ldots$
it is straightforward to compute that%
\begin{align}
c_{q_{m}}^{(1)}(h_{n}^{0}(\alpha)) &  =ig(px^{m-1}+x^{m-1}p)\label{23}\\
c_{q_{m}}^{(2)}(h_{n}^{0}(\alpha)) &  =-4g^{2}x^{2m-2}\label{24}\\
c_{q_{m}}^{(3)}(h_{n}^{0}(\alpha)) &  =0\label{25}%
\end{align}
\noindent for all $n,m\geq0$. With (\ref{23})-(\ref{25}) the generic
expressions (\ref{HHH}) and (\ref{HH}) yield with $\ell=2$
\begin{align}
h_{n,m}^{GS}(\alpha,g) &  =h_{n}^{0}(\alpha)+\frac{1}{2}g^{2}x^{2m-2}%
\label{HS}\\
H_{n,m}^{GS}(\alpha,g) &  =h_{n}^{0}(\alpha)-i\frac{g}{2}(px^{m-1}%
+x^{m-1}p),\label{HGS}%
\end{align}
which are related according to the first relation in (\ref{sim}). In the
special case $n=m=2$, the Hamiltonian $H_{2,2}^{GS}$ becomes the Swanson
Hamiltonian discussed in \cite{Swanson,HJ,Moyal2} upon some change in the
conventions for the coupling constants. This Hamiltonian arises in the second
quantization $H=c_{1}aa+c_{2}a^{\dagger}a^{\dagger}+c_{3}a^{\dagger}a$ where
the $c_{i}$ are coupling constants and $a^{\dagger}=(x-ip)/\sqrt{2}$,
$a=(x+ip)/\sqrt{2}$ are the usual creation and annihilation operators,
respectively. The sequence of Hamiltonians (\ref{HGS}) illustrates our
assertion on the limitations of $\mathcal{PT}$-symmetry in section II A 1,
that there are non-Hermitian Hamiltonians with real energy spectra which are,
however, not $\mathcal{PT}$-symmetric. As one easily sees $H_{n,m}^{GS}%
(\alpha,g)$ is not $\mathcal{PT}$-symmetric when $m$ is odd, but still has a
Hermitian counterpart and therefore real eigenvalues.

Let us next assume that we had simply given the non-Hermitian Hamiltonian and
we wanted to compute the $\eta$-operator. For instance, for $H_{2,2}%
^{GS}(\alpha,g)$ and $H_{4,2}^{GS}(\alpha,g)$ the corresponding equations
(\ref{He}) become%
\begin{align}
0  &  =4gpx\eta^{2}+2\alpha x\partial_{p}\eta^{2}-2p\partial_{x}\eta
^{2}+g\partial_{p}\partial_{x}\eta^{2},\label{65}\\
0  &  =4gpx\eta^{2}+4\alpha x^{3}\partial_{p}\eta^{2}-2p\partial_{x}\eta
^{2}+g\partial_{p}\partial_{x}\eta^{2}-\alpha x\partial_{p}^{3}\eta
^{2},\nonumber
\end{align}
respectively. Both equations are easily solved by $\eta^{2}=\exp(gx^{2})$,
thus confirming our previous calculation.

Having the operator $\eta=\exp(gx^{m}/m)$ at hand we compute from (\ref{obs})
the observables which correspond to the position and momentum operator in the
non-Hermitian systems $H_{n,m}^{GS}(\alpha,g)$ as
\begin{equation}
X=x\qquad\text{and\qquad}P=p-igx^{m-1}, \label{Xx}%
\end{equation}
respectively. Then it is easily verified that indeed (\ref{89}), (\ref{90})
and (\ref{gaugenonherm}) hold.

With regard to the uniqueness of this solution one can see that the first
equation in (\ref{65}) is also solved by $\tilde{\eta}^{2}=\exp(-g/\alpha
p^{2})$. In fact for what has been remarked at the end of the last section, it
is clear that there should be more solutions corresponding to $\check{\eta
}^{2}=\exp(g\hat{x}^{2})f(h_{2}^{0}(\alpha))$, with $f$ being some arbitrary
well behaved function restricted by the demand that $\Phi=\check{\eta}%
^{-1}\phi$ remains a bounded function. Obviously $\check{\eta}^{2}=$
$\tilde{\eta}^{2}$ for $f(x)=\exp(-2gx/\alpha)$. To see that other choices for
$f(x)$ will also lead to solutions of (\ref{65}) is less straightforward as we
have to turn the operator valued expressions for $\check{\eta}^{2}$ first into
real valued functions before we can verify (\ref{65}).

Let us next illustrate how to fix the ambiguities by an explicit choice of the
observables in the non-Hermitian system, which is always possible for what has
been said at the end of section II A. Demanding for instance that\ $X=x$
should be an observable in the non-Hermitian system, it follows immediately
that the only choice for $f(x)$ is $f(x)=1$ and therefore (\ref{Xx}) is the
corresponding set of canonical variables. In turn we could also choose
$\tilde{P}=p$ to be an observable, which leads to $\tilde{\eta}^{2}$ and
$\tilde{X}=x-ig/ap$. For $m\neq n$ it is not possible choose $\tilde{P}=p$ to
be an observable as one can not find a function $f(x)$ such that $\check{\eta
}^{2}$ becomes a function of $p$ only.

\subsection{The spiked Harmonic Oscillator}

A further interesting example is the spiked harmonic oscillator as it exhibits
an explicit supersymmetry \cite{Witten,Cooper,Znojil:1999qt} and therefore
also phenomena like degeneracy of the energy eigenvalues and even level
crossings. The Hermitian version of this Hamiltonian is simply
\begin{equation}
h^{SHO}(x,p)=\frac{1}{2}p^{2}+\lambda^{2}x^{2}+\frac{\alpha^{2}-1/4}{x^{2}}.
\label{Hh}%
\end{equation}
This example is very instructive as it is exactly solvable. The normalized
eigenfunctions are%
\begin{equation}
\phi_{n}^{\alpha}(x)=(-1)^{n}\sqrt{\frac{x\lambda^{\alpha+1}\Gamma
(n+1)}{\Gamma(\alpha+n+1)}}e^{-\frac{\lambda x^{2}}{2}}x^{\alpha}L_{n}%
^{\alpha}(\lambda x^{2}), \label{phia}%
\end{equation}
where the $L_{n}^{\alpha}(x)$ denote the generalized Laguerre polynomials and
the eigenenergies are%
\begin{equation}
\varepsilon_{n}^{\alpha}=\lambda(4n+2\alpha+2).
\end{equation}
Clearly there is a degeneracy of the energy levels for $\varepsilon
_{n}^{-\alpha}=\varepsilon_{n-\alpha}^{\alpha}$. The standard harmonic
oscillator Hamiltonian results from (\ref{Hh}) for $\alpha=\pm1/2$. The
corresponding wavefunctions are related to (\ref{phia}) as $\phi_{n}%
^{1/2}=i^{2n-1}\phi_{2n+1}^{HO}$, $\phi_{n}^{-1/2}=(-1)^{n}\phi_{2n}^{HO}$.
The motivation here to introduce an Hermitian counterpart for this Hamiltonian
is that one wishes to regularize the singularity of the potential at $x=0$,
see e.g. \cite{Znojil:1999qt}.

With $\eta=\exp(-\xi p)$ one easily produces the desired shift and with
(\ref{sim}) one obtains%
\begin{equation}
H^{SHO}(x,p)=\frac{1}{2}p^{2}+\lambda^{2}(x-i\xi)^{2}+\frac{a^{2}-1/4}%
{(x-i\xi)^{2}}. \label{Hxp}%
\end{equation}
This is an example for which the Moyal products are not very suitable for the
computations as the last term in the potential of (\ref{Hxp}) is responsible
for the fact that the related differential equations are of infinite order.

Nonetheless, commutators are easily evaluated in this case and for instance
the canonical variables for the non-Hermitian system are computed in a rather
trivial way, resulting to%
\begin{equation}
X=x-i\xi\qquad\text{and\qquad}P=p. \label{XP}%
\end{equation}
Once again we verify (\ref{89}) and (\ref{90}) for consistency.

\noindent\epsfig{file=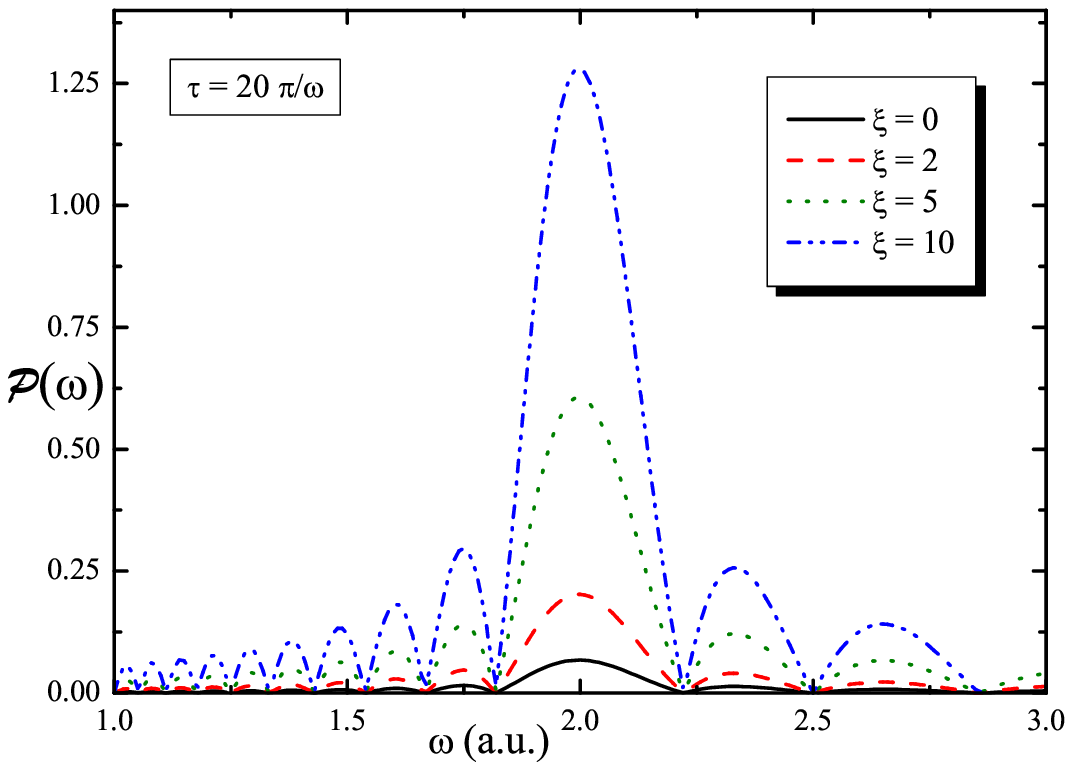,width=9.1cm}

{\small Figure 3: Transition probability for the spiked harmonic oscillator,
as functions of the field frequency }$\omega$ {\small and different parameters
}$\xi$ {\small introduced in (\ref{Hxp}). We consider the transition from the
energy level $n=2$ to $m=3$ to first-order perturbation theory with respect to
the external laser field$.$ The field amplitude is taken to be weak
$E_{0}=0.005$ a.u. and the coupling constant is chosen as $\lambda=0.5$. The
pulse length $\tau$ and the frequency $\omega$ are indicated in the
figure.\medskip}

As this example is completely solvable it serves well to illustrate various
general features. First we use it to argue that adding $xE(t)$ instead of
$XE(t)$ to $H$ in order to construct the non-Hermitian version of the
Stark-LoSurdo Hamiltonian is unphysical. To see this we compute the transition
amplitude in the weak intensity regime to first order (\ref{u1}), where
instead of the amplitude $\left\langle \Phi_{n}\right\vert \!\!\left.
X\Phi_{m}\right\rangle _{\eta}$ we would have $\left\langle \phi
_{n}\right\vert \eta x\eta^{-1}\left.  \phi_{m}\right\rangle $. Now for
(\ref{XP}) we would have that $\left\langle \phi_{n}\right\vert \eta
x\eta^{-1}\left.  \phi_{m}\right\rangle =\left\langle \phi_{n}\right\vert
x\left.  \phi_{m}\right\rangle $ for $n\neq m$, such that no effect would be
visible in the transition amplitude to first order.

Let us therefore take instead the transformation $\tilde{\eta}=\exp(-\xi
p^{2})$. We then compute the canonical variables to%
\begin{equation}
\tilde{X}=x-i2\xi p\qquad\text{and\qquad}\tilde{P}=p.
\end{equation}
For the corresponding non-Hermitian system $H(x,p)=h^{SHO}(\tilde{X},\tilde
{P})$ we evaluate next the transition amplitude for $\phi_{2}^{0.2}%
\rightarrow\phi_{3}^{0.2}$ with $\lambda=0.5$, subjected to a monochromatic
linearly polarized electric field $E(t)=E_{0}\sin(\omega t)$ and depict the
result in figure 3.

As expected we obtain the main contribution for the transition at
$\varepsilon_{3}^{0.2}-\varepsilon_{2}^{0.2}=2$. The value $\xi=0$ is
perfectly reasonable and corresponds to adding $XE(t)$ to $H$, with $X$ given
in (\ref{XP}) for the reasons outlined above. However, for large enough values
of $\xi$ we observe that the transition probability becomes larger than 1,
which is of course inconsistent and unphysical. Therefore to add $xE(t)$ to
$H$ is meaningless in our framework, unless $x$ can be chosen to be an
observable in the non-Hermitian system.

\subsection{The -x$^{4}$ potential}

A further interesting Hamiltonian arises when we specify in equation (\ref{1})
the parameter $N=4$, which involves a potential which is unbounded from below.
Recently Jones and Mateo \cite{JM} established that this Hamiltonian is in
fact isospectral to the Hermitian Hamiltonian%
\begin{equation}
\mathcal{\tilde{H}}=p^{2}+4g^{2}x^{4}-2gx\quad\text{for \ }x,g\in
\mathbb{R}\label{Ht}%
\end{equation}
This Hamiltonian is of great interest as it serves as a simplified version for
the $-\phi^{4}$ quantum field theory, which may for instance be used to mimic
the Higgs mechanism. To obtain the Hamiltonian (\ref{Ht}) from (\ref{1}) with
$N=4$ one needs to pass via two auxiliary Hamiltonians as follows
\begin{equation}
H(z)\overset{z(x)}{\longrightarrow}H(x)\overset{\eta}{\longrightarrow
}h=h^{\dagger}\overset{FT}{\longrightarrow}\mathcal{\tilde{H}}.\label{step}%
\end{equation}
All manipulation in (\ref{step}) are spectrum preserving. In the first step
the general idea \cite{Mostafazadeh:2004tp} was used to map the contour from
within the wedges $\mathcal{W}_{L}$ and $\mathcal{W}_{R}$ back to the real
axis. As discussed in section II A 1 there are many possible parameterization,
which guarantee the appropriate boundary condition. Unfortunately, there is no
constructive method to select out the most useful contour within the wedges
and this choice remains a matter of inspired guess work \cite{JM}. Here the
best choice is guided by the desire to be able to construct a similarity
transformation $\eta$, which maps the non-Hermitian Hamiltonian $H$ adjointly
into a Hermitian Hamiltonian $h$. Hitherto, this procedure was only successful
in an exact manner for the class of Hamiltonians in (\ref{1}) with $N=4$, in
which case $\eta$ can be constructed exactly either by operator methods
\cite{JM}, differential-equation techniques \cite{BBC} or Moyal products
\cite{ACIso}. Even for the next example $N=6$ the same transformation used as
in \cite{JM} does not yield an exact similarity transformation \cite{CAH}. The
last step in (\ref{step}) in the case $N=4$ is to transform $h$ into the
Hamiltonian $\mathcal{\tilde{H}}$ (\ref{Ht}) via a Fourier transformation.

Concretely, we exchange now the constant $g$ by $\varepsilon$ in (\ref{1})
with $N=4$ and obtain $H=-d^{2}/dz^{2}-\varepsilon z^{4}$ thereafter. Using
now the parameterization $z_{1}(x)$ as defined in (\ref{z1}) one obtains the
non-Hermitian Hamiltonian
\begin{equation}
H^{x^{4}}=\hat{p}^{2}-\frac{\hat{p}}{2}+\alpha\hat{x}^{2}-\alpha\,+ig\,\left(
\frac{\{\hat{x},\hat{p}^{2}\}}{2}-2\,\alpha\hat{x}\,\right)  , \label{aq}%
\end{equation}
The domain of $H^{x^{4}}$is now the entire real axis, where $\alpha
=16\varepsilon$ and the coupling constant $g$ has been introduced to separate
off the non-Hermitian part \cite{JM,ACIso}. Next we want to compute $\eta$ by
means of Moyal products. For this we have to convert $H$ first into a real
valued function and have to substitute the anti-commutator with the Moyal
products. Thus we have to replace $\{\hat{x},\hat{p}^{2}\}$ by $x\star
p^{2}+p^{2}\star x=2xp^{2}$. Subsequently we can use (\ref{Moy}) and the
differential equation (\ref{He}) for the Hamiltonian (\ref{aq}) in the unknown
quantity $\eta^{2}(x,p)$ becomes
\begin{align}
0  &  =4gp^{2}x\eta^{2}-8gx\alpha\eta^{2}-4x\alpha\partial_{p}\eta
^{2}\label{d1}\\
&  -\partial_{x}\eta^{2}+4p\partial_{x}\eta^{2}+2gp\partial_{p}\partial
_{x}\eta^{2}-gx\partial_{x}^{2}\eta^{2}.~~\nonumber
\end{align}
We can solve this by
\begin{equation}
\eta^{2}=e^{\frac{g\,p^{3}}{3\,\alpha}-2\,g\,p}, \label{etaa}%
\end{equation}
such that $\eta=e^{\frac{g\,p^{3}}{6\,\alpha}-g\,p}$. From (\ref{sim}) we
obtain thereafter the Hermitian Hamiltonian%
\begin{equation}
h^{x^{4}}=\hat{p}^{2}-\frac{\hat{p}}{2}+\alpha\left(  \hat{x}^{2}-1\right)
\,+g^{2}\frac{\,\left(  \hat{p}^{2}-2\,\alpha\right)  ^{2}}{4\,\alpha}.
\label{d2}%
\end{equation}
Let us compare how these expressions are obtained by means of operator
identities. In principle we have to make a general ansatz to find $q$, but
having already found $\eta$ we can simply extract it from (\ref{etaa})%
\begin{equation}
q=\frac{g}{3\alpha}p^{3}-2gp
\end{equation}
and verify the corresponding expressions. From (\ref{aq}) we find that%
\begin{equation}
h_{0}^{x^{4}}=\hat{p}^{2}-\frac{\hat{p}}{2}+\alpha\left(  \hat{x}%
^{2}-1\right)  . \label{ho}%
\end{equation}
Next we compute the $n$-fold commutators%
\begin{align}
c_{q}^{(1)}(h_{0}^{x^{4}})  &  =4\,ig\alpha x-\,ig\,\{x,p^{2}\}\\
c_{q}^{(2)}(h_{0}^{x^{4}})  &  =-g^{2}\frac{2}{\alpha}\,\left(  p^{2}%
-2\,\alpha\right)  ^{2}\\
c_{q}^{(3)}(h_{0}^{x^{4}})  &  =0.
\end{align}
With $\ell=2$ we then find that the generic expression (\ref{HHH}) for the
Hermitian Hamiltonian yields precisely (\ref{d2}) and the generic expression
(\ref{HH}) for the non-Hermitian Hamiltonian gives (\ref{aq}).

Now the non-Hermitian system in terms of its canonical variables
\begin{equation}
X=x+\frac{ig}{2a}\left(  p^{2}-2a\right)  \qquad\text{and\qquad}P=p,
\end{equation}
results from $H^{x^{4}}(x,p)=h^{x^{4}}(X,P)$. In addition we verify $h^{x^{4}%
}(x,p)=(H^{x^{4}})^{\dagger}(X,P)$.

In this case it suffices to choose $P=p$ as an observable to make the metric
unique. Note also that it is not possible to demand $X=x$ to be an observable
as we can not find a function $f(x)$ such that all functional dependence on
$p$ is eliminated from the term $q+f(h_{0}^{x^{4}})$.

\section{Conclusions}

Given a non-Hermitian time-independent Hamiltonian $H$, we argued that the
analogue of the Stark-LoSurdo Hamiltonian should be
\begin{equation}
H_{l}(t)=H+XE(t),
\end{equation}
where $X=\eta^{-1}x\eta$ is the position operator in the non-Hermitian system.
As we have shown when we simply add $xE(t)$ to $H$, we obtain unphysical
results unless $x$ is an observable in the non-Hermitian system. However, we
also demonstrated that this is not always possible and $x$ is often degraded
to be a mere auxiliary variable in the non-Hermitian system.

As in the time-independent scenario we saw that once the similarity
transformation is known, one can easily translate all the relevant
calculations into the Hermitian system. The situation is less straightforward
when the transformation $\eta$ and therefore the Hermitian system is not
known. In that case one may take our expressions as benchmarks and think of
various different approximation schemes, such as standard perturbation theory,
a perturbation via the $\mathcal{C}$-operator, Floquet type approximations for
periodic potentials etc.

From what has been said one may adopt a rather pessimistic
standpoint and conclude that in the end the non-Hermitian
formulation is in most cases a mere change of metric of a well posed
Hermitian problem. Nonetheless, even leaving the technical
difficulty aside to establish the precise relation between these
conceptually different formulations, it has been successfully argued
that the non-Hermitian formulation is often more natural and
simplifies computations \cite{Bender:2005sc,BBJM}. For an atomic
physicist this is of course a natural scenario when we compare these
alternative formulations with treatments in various gauges, which
are also just different ways to express the same physical quantity.
It is a well established fact that different choices of gauges often
drastically simplify problems in that context and allow for a more
intuitive interpretation. For instance, tunneling processes can be
visualized and interpreted more easily in the length gauge
formulation, since then one may picture the problem in terms of a
time-dependent effective potential barrier, whereas all other gauges
would obscure this intuitive physical interpretation. Furthermore,
phenomena occurring in the context of high frequency fields are most
intuitively understood when viewed in a time-dependent dichotomous
potential in the Kramers-Henneberger gauge

Let us conclude by commenting on some of the immediate open problems, which
follow from what we discussed. Concerning the time-dependent treatment it
would be interesting to change the current set-up by allowing $\eta$ to be time-dependent.

Having entirely focussed on the pseudo-Hermitian nature of the Hamiltonians
involved, we want to conclude with a final comment on the role played by
$\mathcal{PT}$-symmetry in the time-dependent setting. When $[\mathcal{PT}%
,\eta]=0$ the term $XE(t)$ is only $\mathcal{PT}$-symmetric when
$E(-t)=-E(t)$. This means that $\mathcal{PT}$-symmetry depends on the explicit
form of the laser pulse. Taking for instance a typical pulse for a laser field
with frequency $\omega$, amplitude $E_{0}$ and Gaussian enveloping function
$f(t)$, that is of the form $E(t)=E_{0}\sin(\omega t)f(t)$, the term $xE(t)$
would be $\mathcal{PT}$-invariant. However, the perfectly legitimate
replacement $\sin(\omega t)$ $\rightarrow$ $\cos(\omega t)$ in this field
would break the $\mathcal{PT}$-invariance. Recall that in this context the
electric field is treated classically. For a discussion of $\ \mathcal{PT}%
$-symmetry for a full quantum electrodynamic setting we may refer to
\cite{QED1,QED2}. However, for the physical applications we dealt with in this
manuscript, $\mathcal{PT}$-invariance is not a relevant issue, since the pulse
is always chosen such that $H\Phi(0)=\varepsilon\Phi(0)$ and $H\Phi
(\tau)=\varepsilon\Phi(\tau)$. The consequences of \ $\mathcal{PT}$-symmetry
on the eigenvalue problem is therefore only important when considering the
full time-independent eigenvalue problem (\ref{hl}). To investigate this full
solution of (\ref{hl}), the consequences on the non-Hermitian counterpart with
its dressed states \cite{Born} would be extremely interesting \cite{ACPrep}.

\end{document}